# Large upper critical fields and strong coupling superconductivity in the medium-entropy alloy $(Ti_{1/3}Hf_{1/3}Ta_{1/3})_{1-x}Nb_x$

*Longfu Li[1, #], Hongyan Tian[1, #], Xunwu Hu[2,5], Lingyong Zeng[1*], Kuan Li[1], Peifeng Yu[1], Kangwang Wang[1], Rui Chen[1], Zaichen Xiang[1], Dao-Xin Yao[2,3,4], Huixia Luo[1, 3,4*]*

[1] School of Materials Science and Engineering, Sun Yat-sen University, No. 135, Xingang Xi Road, Guangzhou, 510275, P. R. China

[2] School of Physics, Sun Yat-sen University, No. 135, Xingang Xi Road, Guangzhou, 510275, P. R. China

[3] State Key Laboratory of Optoelectronic Materials and Technologies, Sun Yat-sen University, No. 135, Xingang Xi Road, Guangzhou, 510275, P. R. China

[4] Guangdong Provincial Key Laboratory of Magnetoelectric Physics and Devices, Sun Yat-sen University, No. 135, Xingang Xi Road, Guangzhou, 510275, P. R. China

[5] Department of Physics, College of Physics & Optoelectronic Engineering, Jinan University, Guangzhou 510632, China

[#] L. Li and H. Tian contributed equally to this work.

*Corresponding author E-mail: luohx7@mail.sysu.edu.cn (H. Luo) ; zengly57@mail.sysu.edu.cn (L. Zeng)

**Abstract**

Since the discovery of high-entropy superconductors in 2014, superconductivity has remained a focal point of interest in medium- and high-entropy alloys (MEAs-HEAs). Here, we report a series of $(Ti_{1/3}Hf_{1/3}Ta_{1/3})_{1-x}Nb_x$ ($0 \leq x \leq 0.9$) MEA superconductors crystallized in the BCC structure, whose superconductivity was characterized by resistivity, magnetization, and specific heat measurements. The study found that the $(Ti_{1/3}Hf_{1/3}Ta_{1/3})_{1-x}Nb_x$ MEAs exhibit bulk superconductivity. With the doping of Nb, the superconducting transition temperature ($T_c$) increases from 5.31 K to 9.11 K, and the normalized $C_{el}$ jumps at $T_c$, $\Delta C_{el}/\gamma T_c$, and the logarithmically averaged characteristic phonon frequency $\omega_{log}$ exhibit dome-shaped curves. Results from specific heat measurements indicate that the superconductivity is of a strongly coupled *s*-wave type observed at $0 \leq x \leq 0.75$. Furthermore, at low Nb content ($0 \leq x \leq 0.125$), the upper critical field of the samples is larger than the Pauli paramagnetic limit. The strongly coupling behavior and large upper critical field in *s*-wave type $(Ti_{1/3}Hf_{1/3}Ta_{1/3})_{1-x}Nb_x$ MEA superconductors are unusual, as they typically occur in other unconventional superconductors. Thus, $(Ti_{1/3}Hf_{1/3}Ta_{1/3})_{1-x}Nb_x$ may have significant potential in the research and understanding of physical mechanisms.

## I. Introduction

The medium and high-entropy materials typically consist of multi-components and display various unique features (e.g., high-entropy effects in thermodynamics, lattice distortion effects in structure, hysteresis diffusion effects in dynamics, and cocktail effects in performance), resulting in unpredictable properties[1–5]. One of the most attractive properties is their superconductivity. Since 2014, the first high-entropy alloy (HEA) superconductor has been reported[6]. Subsequently, medium and high-entropy materials have emerged as a new platform for investigating superconductivity[7–14]. So far, various medium and high-entropy superconductors have been discovered. Their reliance on valence electron count (VEC) is distinct from that observed in amorphous and crystalline alloy superconductors[15–17], and their superconducting transition temperatures ($T_c$s) also exhibit robustness against high pressure or disorder[18,19]. Additionally, high-entropy carbonitrides possess superconductivity and topological bands, which may be topological superconductors[20–22]. More recently, a TaNbHfZr bulk medium-entropy alloy (MEA) superconductor with a body-centered cubic (BCC) structure shows a dome-shaped superconducting transition temperature ($T_c$) curve with a high-record $T_c \approx 15$ K under 70 GPa[23]. However, the $T_c$ of MEA-HEA superconductors at ambient pressure is generally below 10 K. Therefore, it is still urgently needed to search for high-$T_c$ and high upper-critical field MEA-HEA superconductors.

Previous studies suggest that chemical doping is an effective method to manipulate the composition ratios in MEAs-HEAs, thereby modulating the VEC and $T_c$ of medium and high-entropy superconductors[24–27]; $T_c$ varies with different structures and dopings, indicating the complexity of the potential physical mechanisms in medium and high-entropy superconductors, for which more research is needed. On the other hand, the Nb element plays a vital role in the $T_c$ of the medium and high-entropy superconductors because it is one of the most frequently utilized elements[5]. What is more, doping experiments have demonstrated that MEA and HEA exhibit their highest $T_c$ when VEC $\approx 4.6$ e/a[19,25]. Recently, an extremely strong electron-phonon coupling fully gapped superconductor TiHfNbTa has been reported[13], with a VEC $\approx 4.6$ e/a.

The extremely strong coupling *s*-wave superconductivity in TiHfNbTa MEA is unusual, as it typically occurs in cuprates, pnictides, and other unconventional superconductors. Therefore, we aim to explore the influence of Nb content on superconductivity and its contribution to the superconducting mechanism in the TiHfNbTa MEA superconductor.

In this paper, we present the synthesis of $(Ti_{1/3}Hf_{1/3}Ta_{1/3})_{1-x}Nb_x$ ($0 \leq x \leq 0.9$), crystal structure, experimental study of the interplay between the Nb content and the superconducting properties. XRD results indicate all these compounds have a body-centered cubic (BCC) structure. Superconductivity has been confirmed through temperature-dependent resistivity and magnetization measurements, revealing type-II superconductors. As the Nb content increases, $T_c$ changes from 5.31 K to 9.11 K, and the upper critical field $\mu_0H_{c2}(0)$ exceeds the Pauli paramagnetic limit at low Nb concentrations ($0 \leq x \leq 0.125$), yet the coupling parameter $\Delta C_{el}/\gamma T_c$ changes from 2.24 to 1.65, follows a dome-shaped curve. It can be concluded that with the increase of Nb content, the material transitions change from strongly coupled to extremely strongly coupled and finally to mediumly coupled. Our work explains to some extent what role Nb plays in the superconducting properties of MEA materials.

## II. Experimental Details

Polycrystalline samples of $(Ti_{1/3}Hf_{1/3}Ta_{1/3})_{1-x}Nb_x$ ($0 \leq x \leq 0.9$) were prepared by the arc-melting method, where metal elements (> 99.9% purity) were weighed according to stoichiometric ratios and arc-melted under an argon atmosphere. The crystal structure information of the prepared samples was tested and analyzed by X-ray diffraction (XRD). The XRD patterns were collected using a MiniFlex of Rigaku (Cu Kα radiation) with a step width of 0.01° and a constant scan speed of 5 °/min and a sealed-tube X-ray generator with the Cu target operating at 45 kV and 15 mA. The Rietveld method was employed to refine the raw XRD spectra using the FullProf Suite software package. Surface morphology and elemental distribution were examined using scanning electron microscopy (SEM) coupled with energy-dispersive X-ray spectroscopy (SEM-EDXS). Detailed measurements of electrical transport, specific heat, and magnetization properties were performed using a Physical Property

Measurement System (PPMS-14T, Quantum Design), and the resistivity was measured using the standard four-probe technique.

## III. Results and Discussion

Figure 1(a) presents the bulk XRD patterns of the MEA $(Ti_{1/3}Hf_{1/3}Ta_{1/3})_{1-x}Nb_x$ ($0 \leq x \leq 0.9$). It is clear from the XRD patterns that all the samples are single phases with broad reflections, owing to the high degree of disorder. All compositions crystallize in the cubic BCC structure with the *Im-3m* space group. The right-hand side of figure 1(a) provides detailed diffraction data within the $2\theta$ range of 36 ° to 40 °. The strongest peak in figure 1(a) belonging to the plane (110) shifts to higher angles, which implies the decrease in lattice constant *a* and the compression of the unit cell with increasing *x*. Figure 1(b) shows the values of lattice parameters and unit cell volume. The lattice parameters and unit cell volume exhibit a monotonically decreasing trend with increasing Nb content, which can be attributed to the atomic radius of niobium (146 pm) being smaller than that of titanium (147 pm) and hafnium (159 pm) and equal to that of tantalum (146 pm). The SEM-EDS analysis was conducted to confirm the element distribution and the accurate proportions in the $(Ti_{1/3}Hf_{1/3}Ta_{1/3})_{1-x}Nb_x$ system. As shown in Table S1, the actual element ratios closely match the designed ratios, and uniform elemental distribution was observed, as depicted in figure S1.

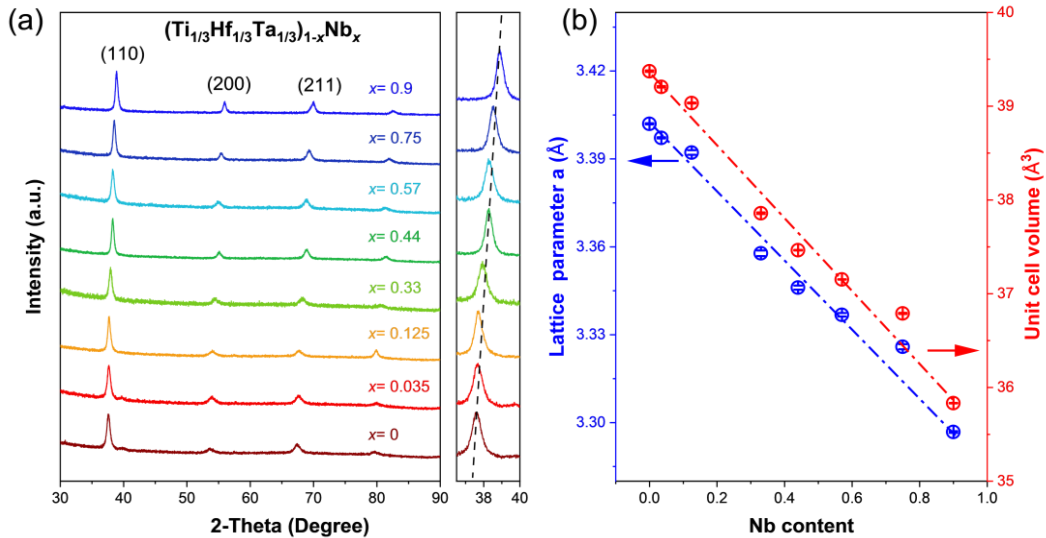

**Figure 1.** (a) XRD patterns of the $(Ti_{1/3}Hf_{1/3}Ta_{1/3})_{1-x}Nb_x$ ($0 \leq x \leq 0.9$) samples. The (h k l) Miller

indices denote the reflections from the BCC structure (space group *Im*-3*m*, No. 229) and diffraction related to the planes (110) in detail. (b) Lattice parameters and unit cell volume as a function of the fraction of Nb in $(Ti_{1/3}Hf_{1/3}Ta_{1/3})_{1-x}Nb_x$ ($0 \leq x \leq 0.9$) samples.

We conducted resistivity measurements to investigate the $T_c$s in the $(Ti_{1/3}Hf_{1/3}Ta_{1/3})_{1-x}Nb_x$ ($0 \leq x \leq 0.9$) compounds. Temperature-dependent resistivity of $(Ti_{1/3}Hf_{1/3}Ta_{1/3})_{1-x}Nb_x$ ($0 \leq x \leq 0.9$) is demonstrated in figure 2(a). The addition of Nb reduces the electrical resistivity. However, considering that mixing entropy first increases and then decreases with increasing $x$ (figure S2), disorder scattering should follow a similar trend. Therefore, the relationship between mixing entropy and resistivity is worth exploring. We have found that in the same doping system, this relationship is not a simple linear one, particularly in highly disordered alloys and complex material systems [28–30]. So, we take the residual resistivity (RRR) as an indicator of changes in the intensity of scattering effect and disorder into account[24,31,32], RRR = $\rho(300K)/\rho(10K)$, the RRR for each sample is exhibited in table 1 and figure S2, are all at relatively low values, comparable to the non-stoichiometric or highly disordered intermetallic compounds observed. Also, the increase in RRR with Nb content points out a diminishing effect in disorder scattering, which might be related to the decreased trend of the reduced electrical resistivity.

To better observe the transition temperatures of different samples, we amplified the temperature-dependent resistance at low temperatures, as shown in figure 2(b). Here, we define the $T_c$ as the 50 % fixed percentage of the normal-state resistance of the sample, with all the samples exhibiting superconductivity. $T_c$ linearly shifts from 5.31 K at $x = 0$ to 9.11 K at $x = 0.9$. Elemental Nb appears to be a crucial element in determining the $T_c$ of highly disordered superconductors, and its presence optimizes $T_c$. In the $Nb_{67}(HfZrTi)_{33}$, Ti-Zr-Hf-Nb-Ta, $(ZrNb)_{1-x}(MoReRu)_x$ and many systems, $T_c$ increases with the Nb content can also be observed [5,11,24,25,33]. In these highly disordered systems, the density of states values is nearly identical under the same electron count and cell volume, a substantial proportion of Nb elements are conducive to elevating the $T_c$ of the BCC-type superconductors. These observations suggest that

Nb is a vital component of intermetallic superconductors. Even at relatively low levels, Nb appears to enhance electron-phonon coupling in some way.

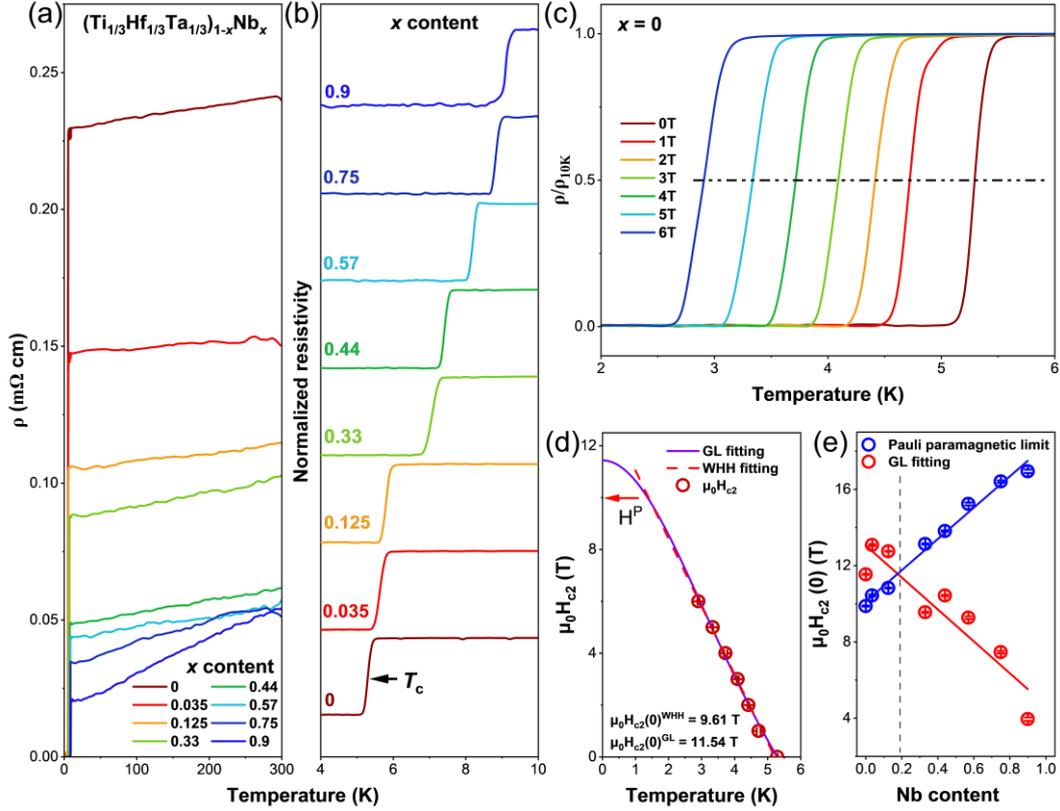

**Figure 2.** (a) Temperature-dependent resistivity of $(Ti_{1/3}Hf_{1/3}Ta_{1/3})_{1-x}Nb_x$ ($0 \leq x \leq 0.9$). (b) The enlarged view of the temperature-dependent resistivity (all curves have been vertically shifted for clarity). (c) The resistivity transition under different magnetic fields. (d) The temperature dependence of the upper critical field $\mu_0H_{c2}$ with the WHH model fitting and the Ginzburg-Landau function fitting. (e) Upper critical fields are a function of the Nb content. The blue line corresponds to the Pauli paramagnetic limit.

By measuring the variation of resistivity with temperature under different applied magnetic fields, the upper critical fields were systematically studied. From figure 2(c), the measured curve of $x = 0$ indicates that $T_c$ decreases and the resistance transition broadens with increasing applied magnetic field. The upper critical field ($\mu_0H_{c2}$) of $x = 0$ is obtained by following the same standards as for zero field resistivity measurements mentioned above through calculations using the Werthamer-Helfand-Hohenburg (WHH) model and Ginzberg-Landau (G-L) theory, which is shown in figure 2(d).

Based on the WHH model, we can calculate the upper critical fields for $x = 0$ as 9.61 T (red dashed line in figure 2(d)). The estimated value of upper critical fields for $x = 0$ is 11.54 T based on the G-L formula: $\mu_0 H_{c2}(T) = \mu_0 H_{c2}(0) \times \frac{1-(T/T_c)^2}{1+(T/T_c)^2}$. According to the Bardeen-Cooper-Schrieffer (BCS) theory, using the equation $H^P = 1.86 \times T_c$. The Pauli limiting field $H^P$ of $x = 0$ is 9.84 T, which is smaller than the values calculated by the GL model. This suggests that the superconductivity may be unconventional and hence indicate that the material may exhibit non-BCS superconductivity.

To quantify the effect of the Nb content on $\mu_0 H_{c2}$, we calculated and summarized the $\mu_0 H_{c2}$ value for all samples for comparison in figure 2(e). The specific test details are shown in figure S3. It can be observed that adding small amounts of Nb improves $\mu_0 H_{c2}$, and upper critical field exceeds the $H^P$ is also observed in $x = 0.035$ and $x = 0.125$, while large amounts of Nb suppress it, even though $T_c$ monotonically increases with Nb content. In some MEA-HEA superconductors, there is partial evidence for the effect of entropy increase on the upper critical field, with $\mu_0 H_{c2}$ generally increasing with entropy [5,24,34], suggesting that strong spin-orbit coupling may play a role in the characteristic properties of the superconducting state in these alloys. Our work appears to further solidify this apparent trend, but more research is needed to confirm it. Additionally, note that the upper critical field is proportional to $\gamma \rho_N$ in the dirty limit, where $\rho_N$ is the resistivity just above $T_c$ in the normal state. Thus, it can be reasonably speculated that the rise in $\rho_N$ and $\gamma$ leads to the $\mu_0 H_{c2}$ being larger than the Pauli paramagnetic limit for $x \leq 0.125$. Furthermore, the changes in the microstructure may influence electron transport and superconducting behavior. For instance, the crystalline and amorphous phases influence the physical properties of high-entropy films. In crystalline/amorphous high-entropy films, superconductivity primarily originates from the nanocrystalline phase, while the amorphous aggregated phase restricts the zero-resistance superconducting state[35,36]. Unusual superconducting transitions are observed only in films with high mixing entropy, indicating that specific compositional ranges significantly affect the collective electronic properties. Superconducting thin films typically exhibit different critical behavior in an external magnetic field, including higher $\mu_0H_{c2}$ [12,37]. Scanning electron microscopy reveals the uniformity of the micro-composition in the current MEA system, and it has also been reported that the BCC structure of HEA superconductors remains uniform at the nanoscale[24]. Therefore, further investigation into how changes in the microstructure of MEA-HEA superconductors affect the stability limit would be very interesting. Consequently, we can calculate the GL coherence length $\xi_{GL}(0)$ using the formulation $\mu_0H_{c2}(0) = \Phi_0/2\pi\xi_{GL}^2$, where $\Phi_0$ represents the quantum flux ($h/2e$). The relevant superconducting parameters of all samples are summarized in Table 1. It can be seen that $\mu_0H_{c2}(0)$ is decreased while the coherence length $\xi_{GL}(0)$ is increased, suggesting that there is a weaker electron-electron interaction at high Nb doping levels.

**Table 1.** Comparison of the superconducting parameters of the MEA $(Ti_{1/3}Hf_{1/3}Ta_{1/3})_{1-x}Nb_x$ ($0 \leq x \leq 0.9$).

| $x$ content | 0 | 0.035 | 0.125 | 0.33 | 0.44 | 0.57 | 0.75 | 0.9 |
|---|---|---|---|---|---|---|---|---|
| $T_c$ (K) | 5.31(1) | 5.61(1) | 5.82(2) | 7.06(6) | 7.41(6) | 8.19(2) | 8.82(5) | 9.11(1) |
| $\mu_0H_{c2}$ (T) | 11.43(3) | 13.08(5) | 12.75(2) | 9.55(3) | 10.43(4) | 9.27(7) | 7.46(6) | 3.95(9) |
| $H^P$ (T) | 9.87(2) | 10.43(3) | 10.82(1) | 13.13(2) | 13.81(2) | 15.23(3) | 16.41(2) | 16.94(4) |
| $\gamma$ (mJ mol$^{-1}$ K$^{-2}$) | 5.935(6) | 5.713(2) | 6.043(2) | 4.838(6) | 5.691(4) | 6.161(3) | 6.840(2) | 8.243(3) |
| $\beta$ (mJ mol$^{-1}$ K$^{-4}$) | 0.393(7) | 0.329(9) | 0.325(3) | 0.250(1) | 0.223(1) | 0.179(4) | 0.134(4) | 0.120(6) |
| $\Theta_D$ (K) | 170.4 | 180.6 | 181.5 | 198.1 | 205.8 | 221.4 | 243.9 | 253.0 |
| $\Delta C_{el}/\gamma T_c$ | 2.20 | 2.26 | 2.47 | 2.70 | 2.43 | 2.36 | 2.01 | 1.65 |
| $\omega_{log}$ (K) | 56 | 58 | 49 | 51 | 65 | 78 | 124 | 245 |
| $\lambda_{ep}$ | 0.77 | 0.78 | 0.79 | 0.84 | 0.83 | 0.85 | 0.84 | 0.84 |
| $2\Delta_0/k_BT_c$ | 4.37 | 4.43 | 4.63 | 4.84 | 4.59 | 4.53 | 4.17 | 3.79 |
| $\xi_{GL}(0)$ (nm) | 5.36 | 5.01 | 5.07 | 5.86 | 5.61 | 5.95 | 6.63 | 9.11 |
| RRR | 1.05 | 1.03 | 1.08 | 1.55 | 1.27 | 1.425 | 1.56 | 2.58 |
| $\Delta S_{mix}$ (R) | 1.09 | 1.20 | 1.33 | 1.35 | 1.29 | 1.15 | 0.82 | 0.43 |

The bulk superconductivity of $(Ti_{1/3}Hf_{1/3}Ta_{1/3})_{1-x}Nb_x$ was characterized through magnetic-susceptibility measurements under an applied field of 3 mT. As shown in figure 3(a), the clear diamagnetic signals were observed below $T_c$ during zero field cooling. The good sample quality is evidenced by the rather sharp transition observed. Considering the demagnetizing factor $N$, the superconducting shielding fraction of all samples is approximately 100 %, indicating bulk superconductivity and 100 % Meissner volume fraction. The $T_c$ determined from the diamagnetic signal is consistent with the electrical resistivity measurements. Additionally, the field-dependent magnetization $M(H)$ curves, as shown in figures 3(c) and 3(d) were collected at various temperatures. The value of magnetization increases linearly and decreases after reaching the lower critical field. We can see that it eventually turns into a paramagnetic state, exhibiting a typical type-II superconductor behavior. Figure 3(b) summarizes the estimated $\mu_0 H_{c1}$ values. Specific test details are shown in figure S4. The result in $\mu_0 H_{c1}$ (0) = 6.5 mT and 34.0 mT for $x = 0$ and $x = 0.57$, respectively.

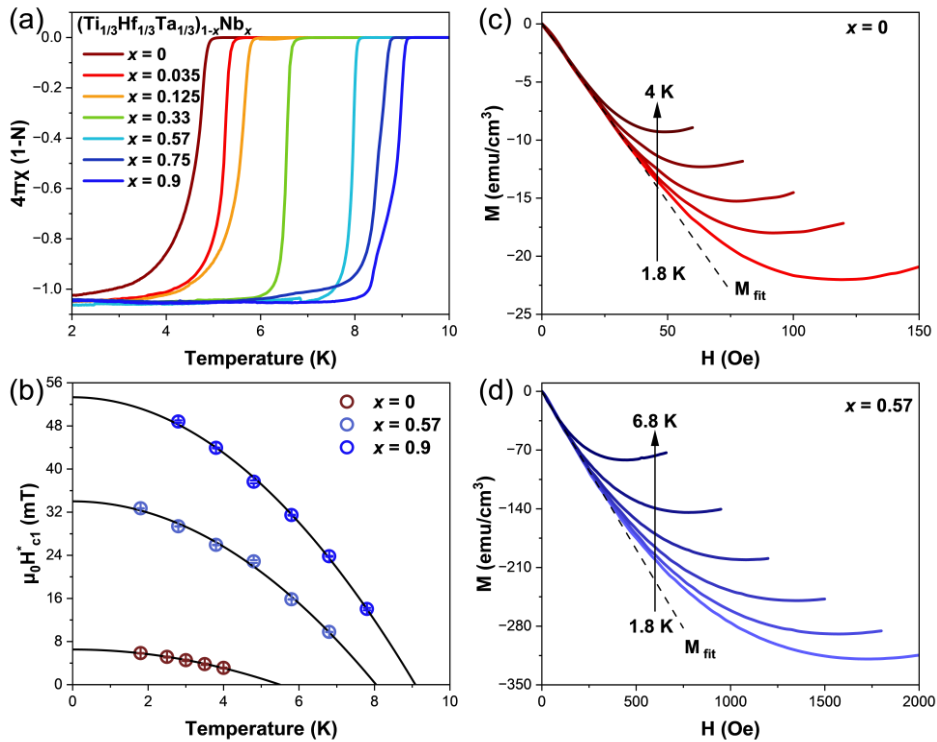

**Figure 3.** (a) The temperature-dependent magnetic susceptibility for $(Ti_{1/3}Hf_{1/3}Ta_{1/3})_{1-x}Nb_x$ ($0 \leq x \leq 0.9$). (b) The temperature dependence of the lower critical fields for $(Ti_{1/3}Hf_{1/3}Ta_{1/3})_{1-x}Nb_x$ ($x = 0, 0.75, 0.9$). Field-dependent magnetization curves were collected at various temperatures

after cooling the (c) TiHfTa and (d) $(Ti_{1/3}Hf_{1/3}Ta_{1/3})_{0.43}Nb_{0.57}$ samples in zero field. For each temperature, $\mu_0H_{c1}$ is determined as the value where $M(H)$ starts deviating from linearity [dashed lines in panels (c) and (d)].

Low-temperature heat capacity measurements were conducted on the MEA $(Ti_{1/3}Hf_{1/3}Ta_{1/3})_{1-x}Nb_x$ samples to confirm their bulk nature of superconductivity and investigate their superconducting pairing. The normal-state specific heat can be fitted using the formula $C_p = \gamma T + \beta T^3$, with $\gamma$ as the normal-state electronic coefficient and $\beta$ as the phonon-specific heat coefficient. We derived the Sommerfeld parameter $\gamma$ = 5.935(6) mJ·mol$^{-1}$·K$^{-2}$ for $x = 0$, while for $x = 0.9$, $\gamma$ = 8.243(3) mJ·mol$^{-1}$·K$^{-2}$. After subtracting the phonon contribution ($\beta T^3$) from the raw data, the electronic specific heat divided by $\gamma$, $C_{el.}/\gamma T$ is obtained. This is shown in figure 4, the normalized heat capacity jumps at $T_c^{mid}$ and $T_c^{zero}$ are estimated to be 2.20 and 1.96 for $x = 0$, respectively. The ratio of $\Delta C_{el}/\gamma T_c$ can reflect the strength of electron coupling. We also calculated the values of $\Delta C_{el}/\gamma T_c$ for other samples (see Table 1), except for $x = 0.9$. All values significantly exceed the BCS weak coupling ratio of 1.43, suggesting that these samples exhibit strong coupling. The fit gives $\beta$ value as 0.393(7) mJ mol$^{-1}$ K$^{-4}$ for $x = 0$ and 0.120(6) mJ mol$^{-1}$ K$^{-4}$ for $x = 0.9$. As $x$ increases, the $\beta$ value decreases continuously, which is also reflected in the rise of the Debye temperature. The Debye temperature can be obtained by $\Theta_D = \left(\frac{12\pi^4 nR}{5\beta}\right)^{1/3}$, where $n$ represents the number of atoms per formula unit, and $R$ is the gas constant. We can obtain the Debye temperatures $\Theta_D$ = 170.4 K and $\Theta_D$ = 253.0 K for $x = 0$ and $x = 0.9$, respectively.

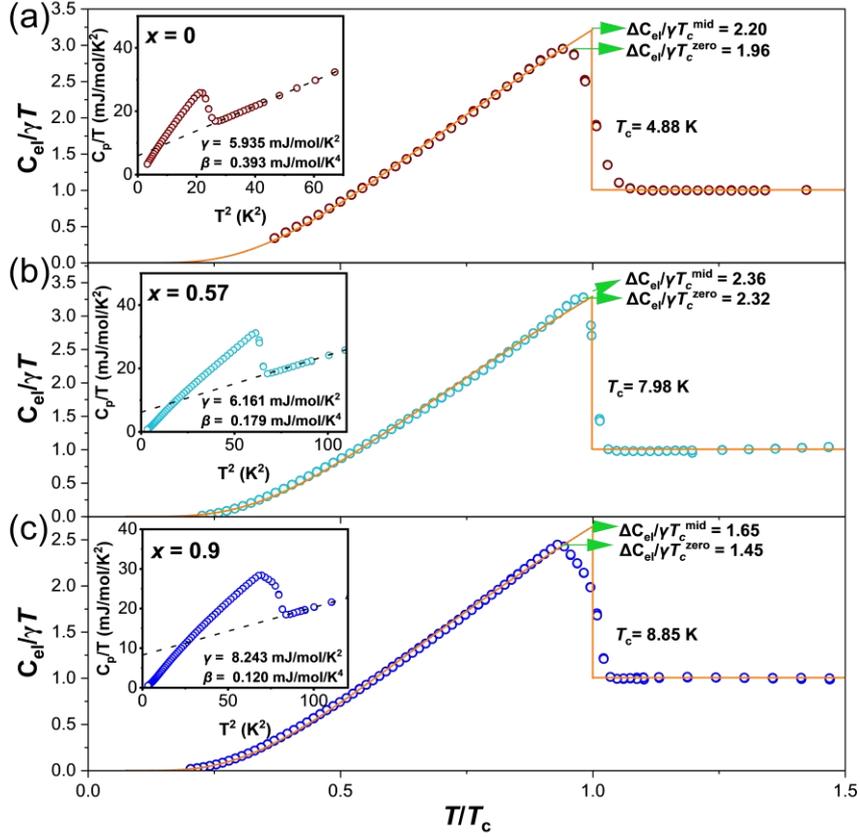

**Figure 4.** Normalized electronic specific heat $C_{el}/\gamma T_c$ as a function of reduced temperature $T/T_c$ for (a) $x = 0$, (b) $x = 0.57$, and (c) $x = 0.9$. The insets show the measured specific heat $C_p/T$ versus $T^2$. The dashed lines in the insets are fit to $C_p = \gamma T + \beta T^3$ for $T > T_c$.

The electron-phonon coupling constant $\lambda_{ep}$ can be calculated from the inverted McMillan formula: $\lambda_{ep} = \dfrac{1.04 + \mu^* \ln\left(\frac{\Theta_D}{1.45 T_c}\right)}{(1-1.62\mu^*)\ln\left(\frac{\Theta_D}{1.45 T_c}\right) - 1.04}$ . The $\mu^*$ represents the coulomb pseudopotential parameter and has the typical value of 0.13[13,38,39]. Their values of $\lambda_{ep}$ are summarized in Table 1 and are close to another strongly coupled superconductor $Ta_{1/6}Nb_{2/6}Hf_{1/6}Zr_{1/6}Ti_{1/6}$[9]. The ratio $\Delta C_{el}/\gamma T_c$ has demonstrated that $(Ti_{1/3}Hf_{1/3}Ta_{1/3})_{1-x}Nb_x$ ($0 \leq x \leq 0.75$) are strongly coupled superconductors. Furthermore, the data can be well described by the conventional s-wave gap function using the α-model, suggesting that the material is fully gapped[40]. The angular independent gap function is expressed as $\Delta(T) = \alpha/\alpha_{BCS}\Delta_{BCS}(T)$ in the α-model, where $\alpha_{BCS}=1.76$ denotes the weak-coupling gap ratio. From the data fitted in figure 4, the superconducting gap values can be obtained, with coupling strength $2\Delta_0/k_B T_c$ being 4.37, 4.43, 4.63, 4.84, 4.59, 4.53 and

4.17 for $x$ = 0, 0.035, 0.125, 0.33, 0.44, 0.57 and 0.75, respectively. Again, these values exceed the weak coupling BCS value (3.52), demonstrating strongly coupled superconductivity in the $(Ti_{1/3}Hf_{1/3}Ta_{1/3})_{1-x}Nb_x$ ($0 \leq x \leq 0.75$) MEAs. The strong coupling s-wave superconductivity is rare. To date, known examples include Pd-Bi alloy[41], pyrochlore osmates[42], and antiperovskite phosphide superconductors[43].

We have plotted the relationship between $\Delta C_{el}/\gamma T_c$ of $(Ti_{1/3}Hf_{1/3}Ta_{1/3})_{1-x}Nb_x$ ($0 \leq x \leq 0.9$) versus Nb content $x$ in figure 5(a). We observed that $\Delta C_{el}/\gamma T_c$ increases in the range of Nb content $x$ from 0 to 0.25. After $\Delta C_{el}/\gamma T_c$ reaches a broad maximum (2.88)[13], then gradually decreases to 1.65 at $x$ = 0.9. According to the modified McMillan equation as corrected by Allen and Dynes[44,45]: $\frac{\Delta C_{el}}{\gamma T_c} = 1.43[1+53(\frac{T_c}{\omega_{log}})^2 \ln(\frac{\omega_{log}}{3T_c})]$, the ratio of $\Delta C_{el}/\gamma T_c$ correlates with the logarithmically averaged characteristic phonon frequency $\omega_{log}$ and $T_c$. Therefore, in figure 5(a), we depicted the evolution of $\omega_{log}$ and $T_c$, where $\omega_{log}$ also shows a dome-like diagram, and $T_c$ is monotonically increasing. So, we can reasonably attribute the dome-like diagram to the varying changes in the logarithmically averaged characteristic phonon frequency $\omega_{log}$ and $T_c$. Figure 5(b) shows the relationship between $2\Delta_0/k_B T_c$ and $\omega_{log}/T_c$. The values were compared in figure 5(b) with those typical strongly coupled superconductors such as Pb and its alloys[41], A15 compounds[41], Chevrel phase compounds[46], β-pyrochlore oxide $KOs_2O_6$[42], and nickel-based $LaO_{0.9}F_{0.1}NiAs$[47]. Each symbol represents a superconductor. The $2\Delta_0/k_B T_c$ parameters of $(Ti_{1/3}Hf_{1/3}Ta_{1/3})_{1-x}Nb_x$ ($0 \leq x \leq 0.9$) fall near the orange line and the trend remains consistent. It is further shown that it can be categorized as a strongly coupled superconductor.

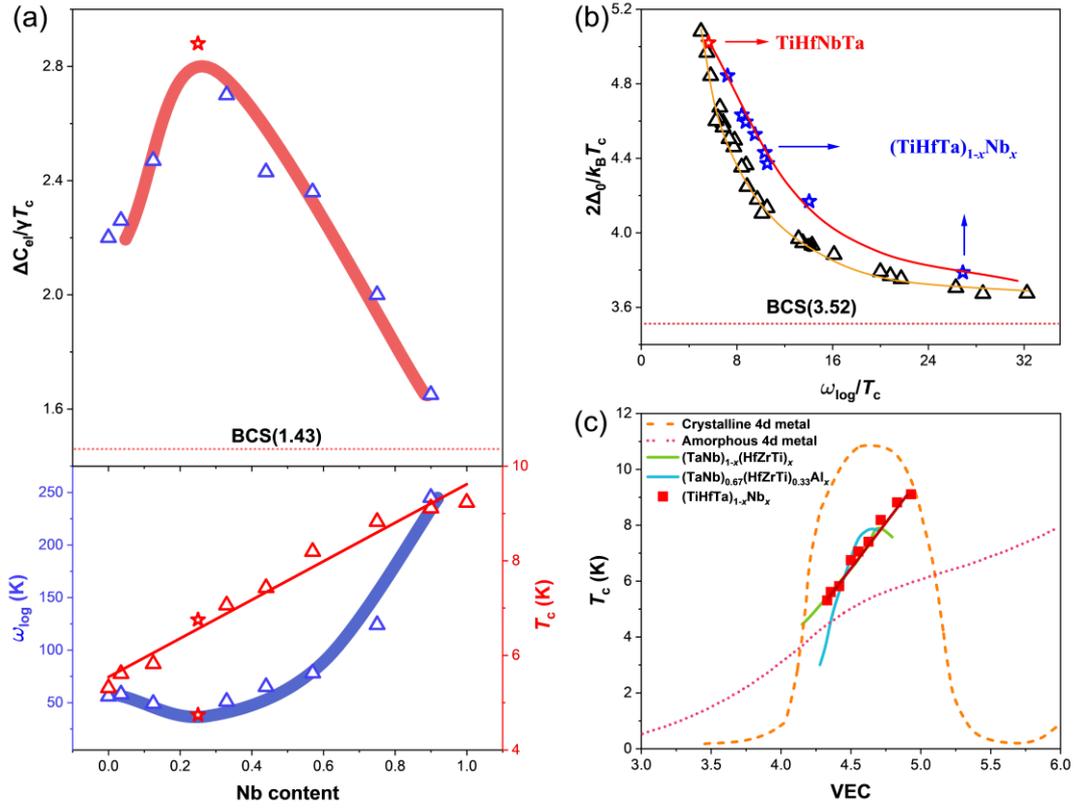

**Figure 5.** (a) The $\Delta C_{el}/\gamma T_c$, the logarithmically averaged characteristic phonon frequency $\omega_{log}$ and $T_c$ under different Nb content. The red star symbol represents the TiHfNbTa sample[13]. (b) The coupling strength $2\Delta_0/k_BT_c$ as a function of the average phonon frequency $\omega_{log}$ normalized by $T_c$. Every symbol is a superconductor (refs. [9,40–42,48]). (c) $T_c$ of $(Ti_{1/3}Hf_{1/3}Ta_{1/3})_{1-x}Nb_x$ ($0 \leq x \leq 0.9$) MEA plotted as a function of the VEC. The data for crystalline 4d metal[49], amorphous 4d metal[50], $(TaNb)_{1-x}(ZrHfTi)_x$[51], and $(TaNb)_{0.67}(ZrHfTi)_{0.33}Al_x$[25] are also included for comparison.

The VEC dependency of the $T_c$ of the $(Ti_{1/3}Hf_{1/3}Ta_{1/3})_{1-x}Nb_x$ ($0 \leq x \leq 0.9$) MEAs are plotted in figure 5(c). The trend lines of the $T_c$ for crystalline forms of transition metals and their alloys, as well as for amorphous vapor-deposited films, are also depicted[49,50]. The trend of transition metals is often referred to as the Matthias rule, which links the $T_c$ maxima with the noninteger d-electron count in simple binary alloys. These two curves are the established standards to which other superconductors may be compared. Studies have shown that the VEC of the MEA or HEA superconductors plays a crucial role in determining their $T_c$ values. Figure 5(c) also shows the relevant data of the $(TaNb)_{1-x}(ZrHfTi)_x$ and $(TaNb)_{0.67}(ZrHfTi)_{0.33}Al_x$ HEA superconductors[25,51]. The observed increase of $T_c$ in $(Ti_{1/3}Hf_{1/3}Ta_{1/3})_{1-x}Nb_x$ ($0 \leq x \leq 0.9$) superconductors is relatively similar to that in crystalline alloys, exhibiting a monotonous increase trend more akin to amorphous superconductors. Additionally, the correlation between $T_c$ and VEC in $(Ti_{1/3}Hf_{1/3}Ta_{1/3})_{1-x}Nb_x$ ($0 \leq x \leq 0.9$) MEAs is highly consistent with that observed in $(TaNb)_{1-x}(ZrHfTi)_x$ HEAs. However, it is important to note that $T_c$ appears to change monotonically with VEC, without reaching a maximum, this may indicate that the Matthias rule has limited applicability to this material, further confirming the unconventional superconductivity of the $(Ti_{1/3}Hf_{1/3}Ta_{1/3})_{1-x}Nb_x$ ($0 \leq x \leq 0.9$) MEAs. Therefore, studying their intrinsic superconductivity helps to understand the differences in superconductivity and superconducting mechanisms between crystalline and amorphous superconductors. The mixing entropy changes across the all chemical compositions used for this study. The largest mixing entropy is present at a ratio of VEC = 4.6 e/a. The $(Ti_{1/3}Hf_{1/3}Ta_{1/3})_{1-x}Nb_x$ ($0 \leq x \leq 0.9$) MEAs can therefore additionally be interpreted as a solid solution ranging from a higher-mixing entropy to a lower one. So, is there a relationship between mixing entropy and superconductivity? For example, whether mixing entropy is a key determinant of $T_c$ in MEA superconductors. By examining Figures 5a and S2, it is evident that the linear variation of $T_c$ does not appear to correlate with the dome-shaped curve of mixing entropy. We have also reviewed several similar studies, and therefore, we can reasonably conclude that increasing mixing entropy does not have a decisive effect on $T_c$ [5,11,24,34,52], but many superconductors with high mixing entropy have very good resistance to high pressure [5,23,53], studying the physical properties of present MEA system under high pressure is also essential. These results suggest that the primary role of high configurational entropy is to stabilize the high-symmetry crystal structure, but has no decisive impact on $T_c$.

## IV. Conclusion

In summary, we have synthesized and characterized a series of $(Ti_{1/3}Hf_{1/3}Ta_{1/3})_{1-x}Nb_x$ ($0 \leq x \leq 0.9$) MEAs with BCC crystal structures, all of which exhibit bulk type-II superconductivity. The $T_c$ is increased from 5.31 K to 9.11 K with increasing $x$. The fitted data show that a higher content of Nb doping reduces the upper critical field of the $(Ti_{1/3}Hf_{1/3}Ta_{1/3})_{1-x}Nb_x$ system. It is worth noting that for $0 \leq x \leq 0.125$, the MEAs with upper critical field larger than the Pauli paramagnetic limit. We speculate that the increase in $\gamma\rho_N$ and entropy leads to the upper critical field being larger than the Pauli paramagnetic limit. Moreover, In the range of $0 \leq x \leq 0.75$, from the high values of $\Delta C_{el}/\gamma T_c$ and $2\Delta_0/k_B T_c$, we demonstrated that the MEAs are fully gapped superconductors with strong electron-phonon coupling. The large upper critical fields and strong coupling superconductivity in the MEAs are unusual. Our research results suggest that further exploration of MEA or HEA superconductors containing Nb would aid in better understanding the physical mechanisms of superconductivity in highly disordered alloy systems.

**Acknowledgments**

This work is funded by the Natural Science Foundation of China (No. 11922415, 12274471, 12404165), Guangdong Basic and Applied Basic Research Foundation (No. 2022A1515011168), Guangzhou Science and Technology Programme (No. 2024A04J6415) and the State Key Laboratory of Optoelectronic Materials and Technologies (Sun Yat-Sen University, No. OEMT-2024-ZRC-02). The experiments and calculations reported were conducted at the Guangdong Provincial Key Laboratory of Magnetoelectric Physics and Devices, No. 2022B1212010008. Lingyong Zeng was thankful for the Postdoctoral Fellowship Program of CPSF (GZC20233299) and Fundamental Research Funds for the Central Universities, Sun Yat-sen University (24qupy092).

# Supporting Information

# Large upper critical fields and strong coupling superconductivity in the medium-entropy alloy $(Ti_{1/3}Hf_{1/3}Ta_{1/3})_{1-x}Nb_x$

*Longfu Li[1,#], Hongyan Tian[1,#], Xunwu Hu[2,5], Lingyong Zeng[1\*], Kuan Li[1], Peifeng Yu[1], Kangwang Wang[1], Rui Chen[1], Zaichen Xiang[1], Dao-Xin Yao[2,3,4], Huixia Luo[1,3,4\*]*

[1] School of Materials Science and Engineering, Sun Yat-sen University, No. 135, Xingang Xi Road, Guangzhou, 510275, P. R. China

[2] School of Physics, Sun Yat-sen University, No. 135, Xingang Xi Road, Guangzhou, 510275, P. R. China

[3] State Key Laboratory of Optoelectronic Materials and Technologies, Sun Yat-sen University, No. 135, Xingang Xi Road, Guangzhou, 510275, P. R. China

[4] Guangdong Provincial Key Laboratory of Magnetoelectric Physics and Devices, Sun Yat-sen University, No. 135, Xingang Xi Road, Guangzhou, 510275, P. R. China

[5] Department of Physics, College of Physics & Optoelectronic Engineering, Jinan University, Guangzhou 510632, China

[#] L. Li and H. Tian contributed equally to this work.

\*Corresponding author E-mail: _luohx7@mail.sysu.edu.cn_ (H. Luo); _zengly57@mail.sysu.edu.cn_ (L. Zeng)

**Table S1.** The element ratios of $(Ti_{1/3}Hf_{1/3}Ta_{1/3})_{1-x}Nb_x$ ($0 \leq x \leq 0.9$) from EDS results.

| Compound | Atom (%) Ti | Hf | Ta | Nb |
|---|---|---|---|---|
| TiHfTa | 31.473 | 36.869 | 31.658 | 0 |
| $(Ti_{1/3}Hf_{1/3}Ta_{1/3})_{0.965}Nb_{0.035}$ | 29.799 | 35.541 | 32.181 | 2.478 |
| $(Ti_{1/3}Hf_{1/3}Ta_{1/3})_{0.875}Nb_{0.125}$ | 27.201 | 34.303 | 27.315 | 11.180 |
| $(Ti_{1/3}Hf_{1/3}Ta_{1/3})_{0.67}Nb_{0.33}$ | 19.475 | 25.506 | 22.961 | 32.058 |
| $(Ti_{1/3}Hf_{1/3}Ta_{1/3})_{0.56}Nb_{0.44}$ | 19.788 | 18.481 | 17.799 | 43.933 |
| $(Ti_{1/3}Hf_{1/3}Ta_{1/3})_{0.43}Nb_{0.57}$ | 13.327 | 15.638 | 13.958 | 57.707 |
| $(Ti_{1/3}Hf_{1/3}Ta_{1/3})_{0.25}Nb_{0.75}$ | 7.748 | 10.430 | 7.401 | 74.421 |
| $(Ti_{1/3}Hf_{1/3}Ta_{1/3})_{0.1}Nb_{0.9}$ | 3.26 | 4.64 | 3.41 | 88.69 |

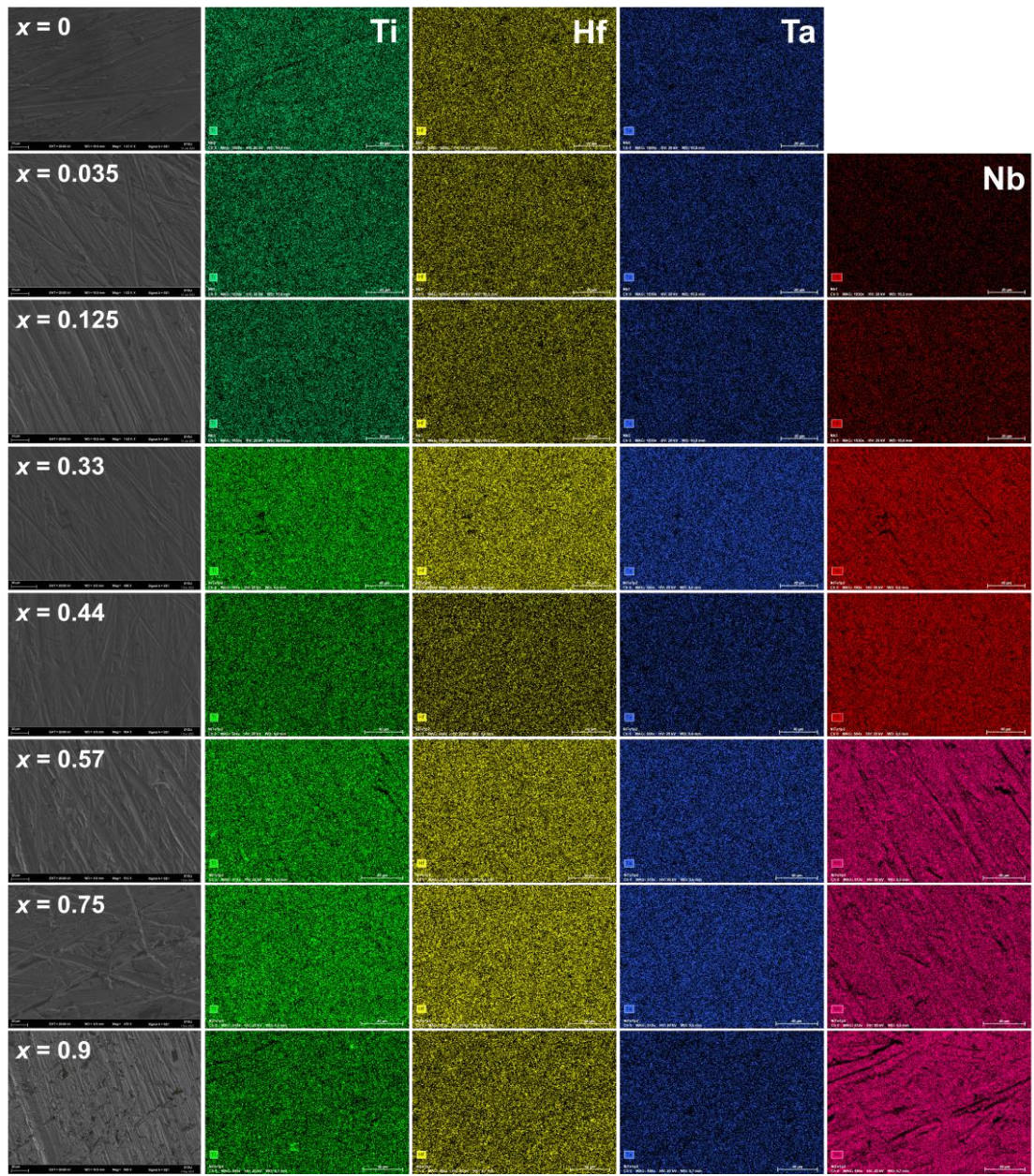

**Figure S1.** SEM images and the corresponding EDS element mappings of $(Ti_{1/3}Hf_{1/3}Ta_{1/3})_{1-x}Nb_x$ ($0 \le x \le 0.9$).

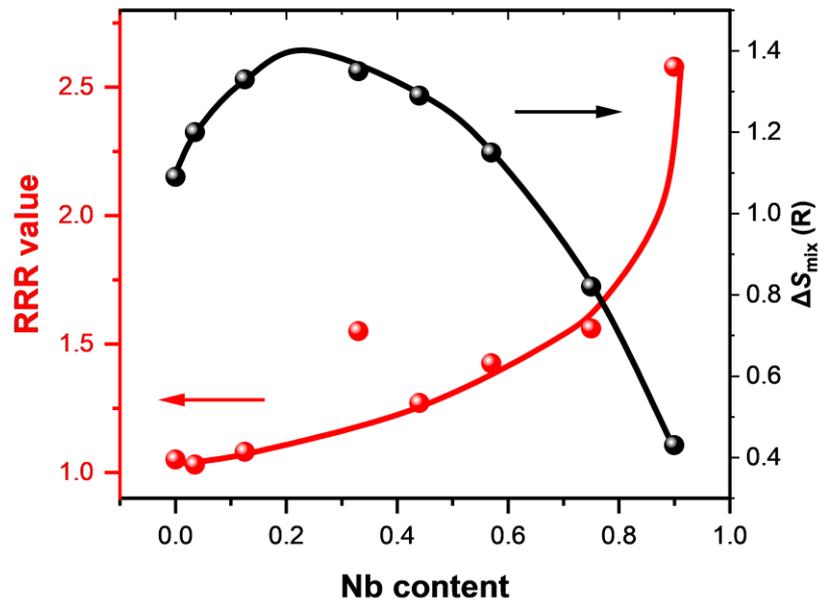

**Figure S2.** The RRR value, and $\Delta S_{mix}$ vs Nb content.

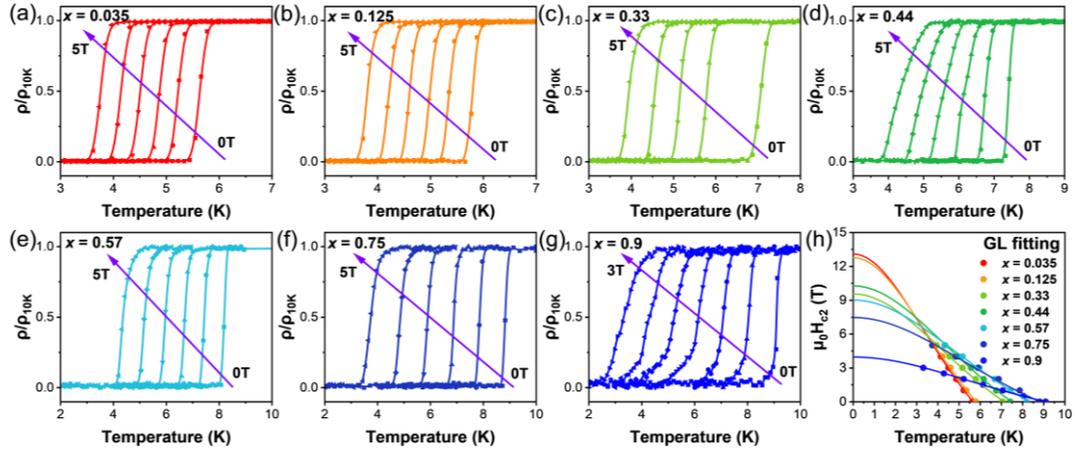

**Figure S3.** (a)-(g) The low-temperature resistivity at different applied fields. (h) The upper critical fields of $(Ti_{1/3}Hf_{1/3}Ta_{1/3})_{1-x}Nb_x$ ($0.035 \leq x \leq 0.9$) samples.

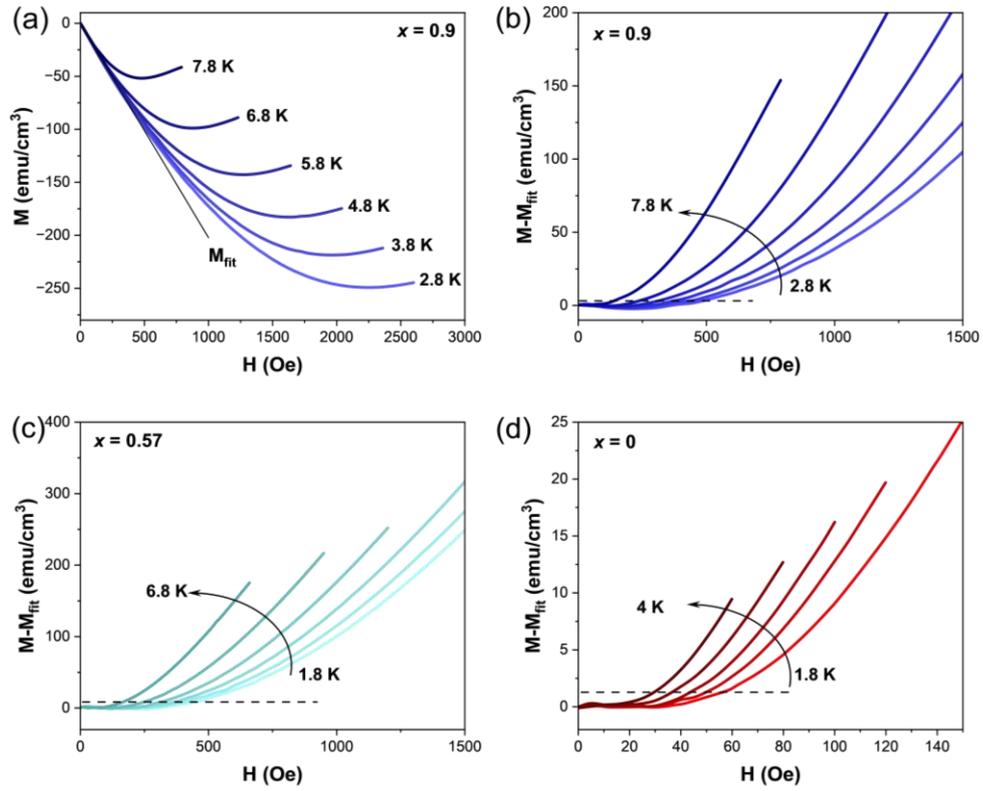

**Figure S4.** (a) The isothermal magnetization curves over a temperature range of 2.8-7.8 K for $x = 0.9$. (b)-(d) The difference between $M$ and $M_{fit}$ under several temperatures for $x = 0$, $x = 0.57$ and $x = 0.9$.

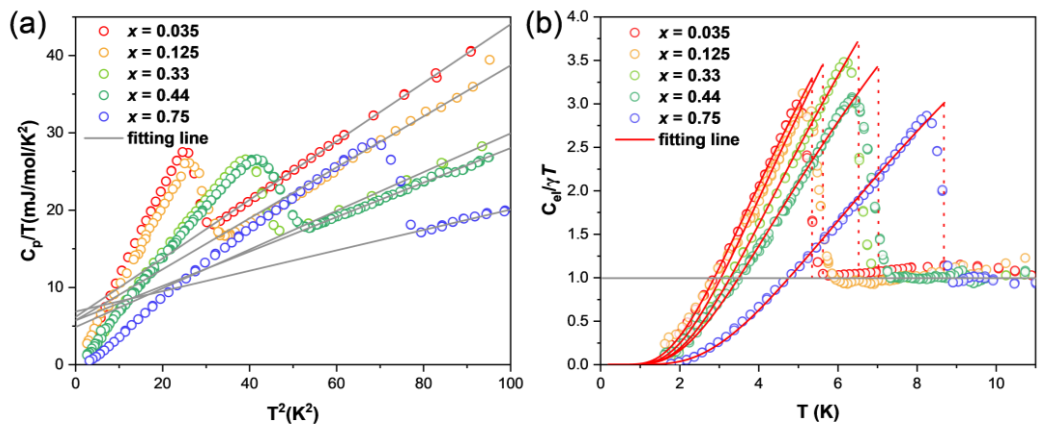

**Figure S5.** (a) $C_p/T$ vs. $T^2$ curves of $(Ti_{1/3}Hf_{1/3}Ta_{1/3})_{1-x}Nb_x$ ($0.035 \leq x \leq 0.75$) samples without magnetic field. (b) the specific heat data fitted with the α-model.